\documentclass[lettersize,journal]{IEEEtran}
\usepackage{orcidlink}
\usepackage{amsmath,amsfonts}
\usepackage{algorithmic}
\usepackage{algorithm}
\usepackage{array}
\usepackage{textcomp}
\usepackage{stfloats}
\usepackage{url}
\usepackage{verbatim}

\usepackage{caption}
\usepackage{subcaption}

\usepackage{graphicx}
\usepackage{cite}
\usepackage{color,soul}
\usepackage{xfrac}

\usepackage{cleveref}
\usepackage{csquotes}
\usepackage{booktabs}
\usepackage{multirow}
\usepackage{tabularx}
\usepackage{microtype}

\usepackage{fancyhdr}
\usepackage{varwidth}
\usepackage{amsmath,amssymb,physics}

\usepackage{tikz}
\usepackage{pgf}
\usepackage{pgfplots}
\usepackage{tikzpagenodes}

\usetikzlibrary{positioning,backgrounds,fit,dsp,calc}
\usetikzlibrary{shapes.geometric}
\usetikzlibrary{shapes.arrows}
\usepackage{array}

\usepackage{xfrac}

\newcommand{\D}[1]{\mathrm{d}{#1}}
\newcommand{\submin}[0]{_\text{min}}
\newcommand{\submax}[0]{_\text{max}}

\newcommand{\T}{\mathrm{T_{60}}}

\DeclareMathOperator*{\argmin}{arg\,min}

\usepackage[nolist, nohyperlinks]{acronym}

\begin{acronym}
\acro{sgm}[SGM]{score-based generative model}
\acro{snr}[SNR]{signal-to-noise ratio}
\acro{gan}[GAN]{generative adversarial network}
\acro{vae}[VAE]{variational autoencoder}
\acro{ddpm}[DDPM]{denoising diffusion probabilistic model}
\acro{stft}[STFT]{short-time Fourier transform}
\acro{istft}[iSTFT]{inverse short-time Fourier transform}
\acro{sde}[SDE]{stochastic differential equation}
\acro{ode}[ODE]{ordinary differential equation}
\acro{ou}[OU]{Ornstein-Uhlenbeck}
\acro{ve}[VE]{Variance Exploding}
\acro{dnn}[DNN]{deep neural network}
\acro{pesq}[PESQ]{Perceptual Evaluation of Speech Quality}
\acro{se}[SE]{speech enhancement}
\acro{tf}[T-F]{time-frequency}
\acro{elbo}[ELBO]{evidence lower bound}
\acro{WPE}{weighted prediction error}
\acro{PSD}{power spectral density}
\acro{RIR}{room impulse response}
\acro{SNR}{signal-to-noise ratio}
\acro{LSTM}{long short-term memory}
\acro{POLQA}{Perceptual Objective Listening Quality Analysis}
\acro{SDR}{signal-to-distortion ratio}
\acro{ESTOI}{Extended Short-Term Objective Intelligibility}
\acro{ELR}{early-to-late reverberation ratio}
\acro{TCN}{temporal convolutional network}
\acro{DRR}{direct-to-reverberant ratio}
\acro{nfe}[NFE]{number of function evaluations}
\acro{rtf}[RTF]{real-time factor}
\acro{mos}[MOS]{mean opinion scores}
\end{acronym}

\begin{document}

\title{Speech Enhancement and Dereverberation with Diffusion-based Generative Models}

\author{Julius Richter\,{\orcidlink{0000-0002-7870-4839}},~\IEEEmembership{Student Member,~IEEE}, Simon Welker\,{\orcidlink{0000-0002-6349-8462}},~\IEEEmembership{Student Member,~IEEE}, Jean-Marie Lemercier\,{\orcidlink{0000-0002-8704-7658}},~\IEEEmembership{Student Member,~IEEE}, Bunlong Lay\,{\orcidlink{0000-0002-0847-7896}}, Timo Gerkmann\,{\orcidlink{0000-0002-8678-4699}},~\IEEEmembership{Senior Member,~IEEE}
\thanks{This work has been funded by the German Research Foundation (DFG) in the transregio project Crossmodal Learning (TRR 169), DASHH (Data Science in Hamburg - HELMHOLTZ Graduate School for the Structure of Matter) with the Grant-No. HIDSS-0002, and the Federal Ministry for Economic Affairs and Climate Action, project 01MK20012S, AP380. We would like to thank J. Berger and Rohde\&Schwarz SwissQual AG for their support with POLQA.}
\thanks{Simon Welker is with the Signal Processing Group, Department of Informatics, Universität Hamburg, 22527 Hamburg Germany, and with the Center for Free-Electron Laser Science, DESY, 22607 Hamburg, Germany (e-mail: simon.welker@uni-hamburg.de). The other authors are all with the Signal Processing Group, Department of Informatics, Universität Hamburg, 22527 Hamburg Germany (e-mail: \{julius.richter; jeanmarie.lemercier; bunlong.lay; timo.gerkmann\}@uni-hamburg.de).}}

\maketitle

\begin{abstract}
In this work, we build upon our previous publication and use diffusion-based generative models for speech enhancement. We present a detailed overview of the diffusion process that is based on a stochastic differential equation and delve into an extensive theoretical examination of its implications. Opposed to usual conditional generation tasks, we do not start the reverse process from pure Gaussian noise but from a mixture of noisy speech and Gaussian noise. This matches our forward process which moves from clean speech to noisy speech by including a drift term. We show that this procedure enables using only 30 diffusion steps to generate high-quality clean speech estimates. By adapting the network architecture, we are able to significantly improve the speech enhancement performance, indicating that the network, rather than the formalism, was the main limitation of our original approach. In an extensive cross-dataset evaluation, we show that the improved method can compete with recent discriminative models and achieves better generalization when evaluating on a different corpus than used for training. We complement the results with an instrumental evaluation using real-world noisy recordings and a listening experiment, in which our proposed method is rated best. Examining different sampler configurations for solving the reverse process allows us to balance the performance and computational speed of the proposed method. Moreover, we show that the proposed method is also suitable for dereverberation and thus not limited to additive background noise removal. Code and audio examples are available online\footnote{\url{https://github.com/sp-uhh/sgmse}}.
\end{abstract}


\section{Introduction}

\IEEEPARstart{S}{peech} enhancement aims to recover clean speech signals from audio recordings that are impacted by acoustic noise or reverberation~\cite{hendriks2013dft}. To this end, computational approaches often exploit the different statistical properties of the target and interference signals~\cite{gerkmann2018book_chapter}. Machine learning algorithms can be used to extract these statistical properties by learning useful representations from large datasets. A wide class of methods employed for speech enhancement are discriminative models that learn to directly map noisy speech to the corresponding clean speech target \cite{wang2018supervised}. Common approaches include \ac{tf} masking \cite{williamson2015complex}, complex spectral mapping \cite{fu2017complex}, or operating directly in the time domain \cite{fu2017raw}. These supervised methods are trained with a variety of clean/noisy speech pairs containing multiple speakers, different noise types, and a large range of \acp{snr}. However, it is nearly impossible to cover all possible acoustic conditions in the training data to guarantee generalization. Furthermore, some discriminative approaches have been shown to result in unpleasant speech distortions that outweigh the benefits of noise reduction~\cite{wang2019bridging}.

The use of generative models for speech enhancement, on the other hand, follows a different paradigm, namely to learn a prior distribution over clean speech data. Thus, they aim at learning the inherent properties of speech, such as its spectral and temporal structure. This prior knowledge can be used to make inferences about clean speech given noisy or reverberant input signals that are assumed to lie outside the learned distribution. Several approaches follow this idea and utilized deep generative models for speech enhancement \cite{pascual2017segan, bando2018statistical, leglaive2018variance, richter2020speech, carbajal2021guided, carbajal2021disentanglement, bando2020adaptive, fang2021variational, nugraha2020flow, bie2022unsupervised}. Among them are methods that employ likelihood-based models for explicit density estimation such as the \ac{vae} \cite{kingma2014auto}, or leverage  \acp{gan} \cite{goodfellow2014generative} for implicit density estimation. Bando et al. propose a statistical framework using a VAE trained in an unsupervised fashion to learn a prior distribution over clean speech \cite{bando2018statistical}. At test time they combine the speech model with a low-rank noise model to infer the signal variances of speech and noise to build a Wiener filter for denoising. However, since the VAE is trained with clean speech only, the inference model (i.e. the encoder) that predicts the latent variable remains sensitive to noise. This has been shown to cause the generative speech enhancement method to produce speech-like sounds although only noise is present \cite{bando2018statistical}. To mitigate this, it has been proposed to make the inference model robust to noisy speech by training on labeled data in a supervised manner \cite{bando2020adaptive, fang2021variational}, or by disentangling the latent variable from high-level information such as speech activity which can be estimated by supervised classifiers \cite{carbajal2021guided, carbajal2021disentanglement}. Nevertheless, VAE-based speech enhancement methods remain limited due to the dimensionality reduction in the latent layer and the combined use of a linear noise model based on non-negative matrix factorization \cite{bando2018statistical, leglaive2018variance, richter2020speech, carbajal2021guided, carbajal2021disentanglement, bando2020adaptive, fang2021variational}.

More recently, a new class of generative models called diffusion-based generative models, has been introduced to the task of speech enhancement \cite{lu2021study, lu2022conditional, welker2022speech, serra2022universal}. Diffusion-based generative models, or simply \emph{diffusion models}, are inspired by non-equilibrium thermodynamics and exist in several variants \cite{sohl2015deep,ho2020denoising,song2019generative}. All of them share the idea of gradually turning data into noise, and training a neural network that learns to invert this process for different noise scales. More specifically, the inference model is a fixed Markov chain, that slowly transforms the data into a tractable prior, such as the standard normal distribution. The generative model is another Markov chain that is trained to revert this process iteratively \cite{ho2020denoising}. Therefore, diffusion models can be considered as deep latent variable models and have similar properties to \acp{vae}, with the crucial difference that the inference model is not trained and that the latent variables have the same dimensionality as the input. This has the advantage of not relying on surrogate objectives to approximate maximum likelihood training such as the evidence lower bound and enforces no strong restrictions on the model architecture. Recently, diffusion models have been connected with score matching \cite{hyvarinen2005estimation} by looking at the \ac{sde} associated with the discrete-time Markov chain \cite{song2021sde}. The forward process can be inverted, resulting in a corresponding reverse \ac{sde} which depends only on the score function of the perturbed data \cite{anderson1982reverse}. Using this continuous-time \ac{sde} formalism creates the opportunity to design novel diffusion processes that support the underlying generation task. In contrast to discrete Markov chains, it also allows the use of general-purpose \ac{sde} solvers to numerically integrate the reverse process for sampling.

Concerning the application of diffusion models for speech enhancement, there exist currently two approaches that differ conceptually in how the diffusion process is used. One approach is based on speech re-generation, i.e. a diffusion-based vocoder network is used to synthesize clean speech by sampling from an unconditional prior, while a conditioner network takes noisy speech as input and performs the core part of denoising by providing enhanced speech representations to the vocoder network \cite{koizumi2022specgrad, serra2022universal}. An auxiliary loss is introduced for the conditioner network to facilitate its ability to estimate clean speech representations \cite{serra2022universal}. The second approach, on the other hand, does not require any auxiliary loss and is not using two separate models for generation and denoising. Instead, it  models the corruption of clean speech by environmental background noise or reverberation directly within the forward diffusion process, so that reversing this process would consequently result in generating clean speech. This has been proposed as a discrete diffusion process for time-domain speech signals \cite{lu2022conditional}, and as a continuous \ac{sde}-based diffusion process in the complex spectrogram domain \cite{welker2022speech}. Interestingly, the original denoising score matching objective \cite{vincent2011connection}, which is to estimate the white Gaussian noise in the perturbed data, is essentially reminiscent of the goal of speech enhancement, which is to remove interfering noise or reverberation from speech signals. However, under realistic conditions, the environmental noise or reverberation may not match the assumption of stationary white Gaussian noise. Therefore, it was proposed to include real noise recordings in the diffusion process, either by linearly interpolating between clean and noisy speech along the process \cite{lu2022conditional}, or by defining such a transformation within the drift term of an \ac{sde} \cite{welker2022speech}. The choice of linear interpolation in \cite{lu2022conditional}, however, implies that the trained \ac{dnn} must explicitly estimate a portion of environmental noise at each step in the reverse process. This can be seen in the resulting objective function \cite[Eq. (21)]{lu2022conditional} which exhibits characteristics of a discriminative learning task. In contrast, an \ac{sde}-based formulation results in a pure generative objective function \cite[Eq. (9)]{welker2022speech} and avoids any prior assumptions on the noise distribution.

Nonetheless, note that diffusion-based speech enhancement methods, unlike the VAE-based method described above, are not counted as unsupervised methods, since labeled data  (i.e. clean and noisy speech pairs) are used for training. However, the learning objective remains generative in nature which is to learn a prior for clean speech per se rather than a direct mapping from noisy to clean speech. In fact, supervision is only exploited to learn the conditional generation of clean speech when noisy speech is given. Thus, current diffusion-based models for speech enhancement, such as \cite{serra2022universal,lu2022conditional,welker2022speech}, can be considered as conditional generative models trained in a supervised manner. 

In this work, we build upon our previous publication which defines the diffusion process in the complex \ac{stft} domain \cite{welker2022speech}. We present a comprehensive theoretical review of the underlying score-based generative model and include an expanded discussion on the conditional generation process which is based on the continuous-time \ac{sde} formalism. By using a network architecture developed in the image processing community \cite{song2021sde}, in the work at hand we significantly improve performance in comparison to our previous model \cite{welker2022speech}. This indicates that the network, rather than the formalism, was the main limitation of our original approach. In an extensive cross-dataset evaluation, we show that the improved method can compete with recent discriminative models and achieves better generalization when evaluating on a different corpus than used for training. To confirm the effectiveness of the proposed method on non-simulated data, we perform an instrumental evaluation with real-world noisy recordings using non-intrusive metrics. 
We complement the results with a listening experiment, in which our proposed method is rated best. Interestingly, using the improved network, we show that the proposed method is also suitable for dereverberation when an individual model is trained on simulated reverberant data. Thus, the method is not limited to the removal of additive background noise and can also be applied to non-additive corruptions such as reverberation or, as shown in \cite{lemercier2023analysing}, for bandwidth extension. Furthermore, we investigate different sampler configurations for solving the reverse process which reveals a trade-off between the performance and computational speed of the proposed method. 

We summarize our major contributions as follows. Regarding the novelty with respect to Song et al. \cite{song2021sde}, we introduce a drift term to the \ac{sde} to achieve the required task adaptation for reconstruction problems and furthermore apply the diffusion process and score matching objective to a complex data representation. Also note that the approach in \cite{song2021sde} is not explicitly trained on reconstruction tasks and the application is different from ours. 
Regarding the novelty with respect to our previous publication \cite{welker2022speech},
we use an improved network architecture and increase the performance significantly. Moreover, we include an extended theoretical discussion and investigate different sampler configurations. Finally, we expand the evaluation by means of a cross-dataset evaluation, an instrumental evaluation with real-world noisy recordings, and a listening experiment.

\section{Method: Score-based Generative Model for Speech Enhancement (SGMSE)}

In this section, we motivate and describe in detail the approach of using score-based generative models for speech enhancement, as proposed in our previous publication~\cite{welker2022speech}.

\subsection{Data representation}

We represent our data in the complex-valued \ac{stft} domain, as it has been observed that both real and imaginary parts of clean speech spectrograms exhibit clear structure and are therefore amenable to deep learning models \cite{williamson2015complex}. Following the approach of complex spectral mapping \cite{fu2017complex}, we use our conditional generative model to estimate the clean real and imaginary spectrograms from the noisy ones. 

The use of complex coefficients as data representation allows the definition of the diffusion process in the complex spectral domain, in which additive Gaussian noise corresponds to the signal model used for the denoising task. This relates to traditional STFT-based methods, where spectral coefficients are usually assumed to be complex Gaussian distributed and mutually independent \cite{hendriks2013dft, gerkmann2018book_chapter}. Statistical approaches often consider an additive signal model assuming that the speech process and the noise process are realizations of stochastic processes that are statistically independent. Observing that the overall noise process is a sum of several independent sources, the central limit theorem ensures that the observed noise process tends to be Gaussian \cite{hendriks2013dft}.

Although it would be theoretically possible to define the diffusion process in the magnitude domain, additive Gaussian noise would not relate to the signal model anymore. This becomes evident considering that in the magnitude domain, additive Gaussian noise could result in negative amplitudes which are physically not defined.

Thus, we operate on complex spectrograms that are elements of $\mathbb{C}^{K \times F}$, where $K$ denotes the number of time frames dependent on the audio length, and $F$ represents the number of frequency bins. 
To compensate for the typically heavy-tailed distribution of \ac{stft} speech amplitudes~\cite{gerkmann2010empirical}, we apply an amplitude transformation 
\begin{equation}\label{eq:spec-transform}
    \tilde{c} = \beta |c|^\alpha \mathrm e^{i \angle(c)} 
\end{equation}
to all complex \ac{stft} coefficients $c$, where $\angle(\cdot)$ represents the angle of a complex number, $\alpha \in (0, 1]$ is a compression exponent which brings out frequency components with lower energy (e.g. fricative sounds of unvoiced speech)~\cite{braun2021consolidated}, and $\beta \in \mathbb R_+$ is a simple scaling factor to normalize amplitudes roughly to within $[0,1]$. Such a compression has been argued to be perceptually more meaningful in speech enhancement \cite{you2005spl, breithaupt2010analysis}, and the transformation ensures that the neural network operates on consistently scaled inputs with respect to the Gaussian diffusion noise \cite{ho2020denoising}.

\subsection{Stochastic Process}

The tasks at hand, speech enhancement and dereverberation, can be considered as conditional generation tasks: Given the corrupted noisy/reverberant speech, generate clean speech by using a conditional generative model. Most previously published diffusion-based generative models are adapted to such conditional tasks either through explicit conditioning channels added to the \ac{dnn}~\cite{chen2021wavegrad,batzolis2021conditional}, or through combining an unconditionally trained score model with a separate model (such as a classifier) that provides conditioning in the form of a gradient~\cite{dhariwal2021diffusion,song2021sde}. With our method, we explore a third possibility, which is to incorporate the particular task directly into the forward and reverse processes of a diffusion-based generative model.

\paragraph{Forward Process} Following Song et al. \cite{song2021sde}, we design a stochastic diffusion process $\{\mathbf x_t\}_{t=0}^T$ that is modeled as the solution to a linear \ac{sde} of the general form,
\begin{equation} \label{eq:ouve-sde}
    \D{\mathbf x_t} = \mathbf f(\mathbf x_t, \mathbf y) \D{t} + g(t) \D{\mathbf w}\,,
\end{equation}
where $\mathbf x_t$ is the current process state, $t \in [0, T]$ a continuous time-step variable describing the progress of the process (not to be confused with the time index of any signal in the time or \ac{tf} domain), $\mathbf y$ the noisy or reverberant speech, and $\mathbf w$ denotes a standard Wiener process. The vector-valued function $\mathbf f(\mathbf x_t, \mathbf y)$ is referred to as the \emph{drift coefficient}, while $g(t)$ is called the \emph{diffusion coefficient} and controls the amount of Gaussian white noise injected at each time-step. 
Note that different to Song et al. \cite{song2021sde}, our drift term is now a function of $\mathbf y$, by which we tailor the proposed \ac{sde} to reconstruction tasks.
The process is defined for each \ac{tf} bin independently. Thus, the variables in bold are assumed to be vectors in $\mathbb C^d$ with $d=KF$ containing the coefficients of a flattened complex spectrogram.

\begin{figure}[t]
\centering
\includegraphics[width=\columnwidth]{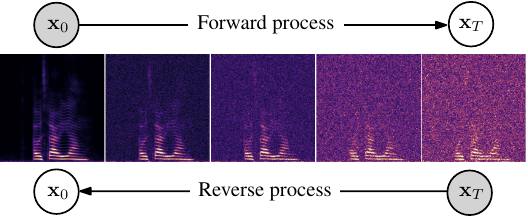}
\caption{Diffusion process on a spectrogram: In the forward process noise is gradually added to the clean speech spectrogram $\mathbf x_0$, while the reverse process learns to generate clean speech in an iterative fashion starting from the corrupted signal $\mathbf x_T$.}
\label{fig:sde-process}
\end{figure}

The forward process in Eq. \eqref{eq:ouve-sde} turns a clean speech sample $\mathbf x_0$ into a corrupted sample $\mathbf x_T$ by gradually adding noise from the Wiener process, as illustrated in Fig. \ref{fig:sde-process}. To account for the intended task adaptation of speech enhancement or dereverberation, we propose a drift term that ensures the mean of the process moving from clean speech $\mathbf x_0$ to noisy/reverberant speech $\mathbf y$. In particular, we define the drift coefficient $\mathbf f$ and the diffusion coefficient $g$ as
\begin{equation}\label{eq:drift_term}
\mathbf f(\mathbf x_t, \mathbf y) := \gamma(\mathbf y-\mathbf x_t)\,,
\end{equation}\vspace{-0.5em}
\begin{equation}
g(t) :=  \sigma\submin \left(\frac{\sigma\submax}{\sigma\submin}\right)^t \sqrt{2\log\left(\frac{\sigma\submax}{\sigma\submin}\right)}\,, 
\end{equation}
where $\gamma$ is a constant called \emph{stiffness} controlling the transition from $\mathbf x_0$ to $\mathbf y$, and $\sigma\submin$ and $\sigma\submax$ are parameters defining the noise schedule of the Wiener process. 
Note that we choose the diffusion coefficient identical to that of the so-called \emph{Variance Exploding \ac{sde}} from Song et al.~\cite{song2021sde}. Our novel contribution lies in the modified drift term, by which the intended task adaptation is achieved.

\paragraph{Reverse Process} Following Anderson~\cite{anderson1982reverse} and Song et al.~\cite{song2021sde}, the SDE in Eq. \eqref{eq:ouve-sde} has an associated \emph{reverse \ac{sde}},
\begin{equation}\label{eq:reverse-sde}
   \hspace*{-0.19em} \D{\mathbf x_t} =
        \left[
            \mathbf f(\mathbf x_t, \mathbf y) - g(t)^2\nabla_{\mathbf x_t} \log p_t(\mathbf x_t|\mathbf y)
        \right] \D{t}
        + g(t)\D{\bar{\mathbf w}}\,,
\end{equation}
where the \emph{score} $\nabla_{\mathbf x_t} \log p_t(\mathbf x_t|\mathbf y)$ is the term to be approximated by a \ac{dnn} which is therefore called a \emph{score model}. We denote the score model as $\mathbf s_\theta(\mathbf x_t, \mathbf y, t)$, which is parameterized by a set of parameters $\theta$ and receives the current process state $\mathbf x_t$, the noisy speech $\mathbf y$, and the current time-step $t$ as an input. Finally, by substituting the score model into the reverse \ac{sde} in Eq.~\eqref{eq:reverse-sde}, we obtain the so-called \emph{plug-in reverse \ac{sde}} \cite{huang2021variational},
\begin{equation}\label{eq:plug-in-reverse-sde}
    \D{\mathbf x_t} =
        \left[
            \mathbf f(\mathbf x_t, \mathbf y) - g(t)^2\mathbf s_\theta(\mathbf x_t, \mathbf y, t)
        \right] \D{t}
        + g(t)\D{\bar{\mathbf w}}\,,
\end{equation}
which can be solved by various solver procedures, to be discussed in detail in Sec. \ref{sec:sampling}.

For inference, we assume that a trained score model $\mathbf s_\theta$ is given, which approximates the true score for all $t\in [0, T]$. We can then generate clean speech $\mathbf x_0$ conditioned on the noisy or reverberant speech $\mathbf y$ by solving the plug-in reverse \ac{sde} in Eq. \eqref{eq:plug-in-reverse-sde}. To determine the initial condition of the reverse process at  $t=T$, we sample
\begin{equation}
\label{eq:init_dist_reverse_process}
    \mathbf x_T \sim \mathcal N_{\mathbb C}(\mathbf x_T; \mathbf y, \sigma(T)^2 \mathbf I), 
\end{equation}
which is a strongly corrupted version of the noisy speech $\mathbf y$, as illustrated in Fig. \ref{fig:sde-process}. The denoising process which solves the task of speech enhancement or dereverberation is then based on iterating through the reverse process starting at $t=T$ and ending at $t=0$.

\subsection{Training objective}

Next, we derive the objective function used for training the score model $\mathbf s_\theta$. Since the \ac{sde} in Eq. \eqref{eq:ouve-sde} describes a Gaussian process, the mean and variance of the process state $\mathbf x_t$ can be derived when its initial conditions are known \cite{sarkka2019sde}. This allows for direct sampling of $\mathbf x_t$ at an arbitrary time step $t$ given $\mathbf x_0$ and $\mathbf y$ by using the so-called \emph{perturbation kernel},
\begin{equation}
\label{eq:perturbation-kernel}
    p_{0t}(\mathbf x_t|\mathbf x_0, \mathbf y) = \mathcal{N}_\mathbb{C}\left(\mathbf x_t; \boldsymbol \mu(\mathbf x_0, \mathbf y, t), \sigma(t)^2 \mathbf{I}\right),
\end{equation}
where $\mathcal{N}_\mathbb{C}$ denotes the circularly-symmetric complex normal distribution and  $\mathbf I$ denotes the identity matrix. We utilize Eqs. (5.50, 5.53) in Särkkä~\&~Solin~\cite{sarkka2019sde} to determine closed-form solutions for the mean
\begin{equation}
\label{eq:mean}
    \boldsymbol\mu(\mathbf x_0,\mathbf y, t) = \mathrm e^{-\gamma t} \mathbf x_0 + (1-\mathrm e^{-\gamma t}) \mathbf y
    \,,
\end{equation}
and the variance
\begin{equation}
    \label{eq:std}
    \sigma(t)^2 = \frac{
        \sigma\submin^2\left(\left(\sfrac{\sigma\submax}{\sigma\submin}\right)^{2t} - \mathrm e^{-2\gamma t}\right)\log(\sfrac{\sigma\submax}{\sigma\submin})
    }{\gamma+\log(\sfrac{\sigma\submax}{\sigma\submin})}
    \,.
\end{equation}

Vincent \cite{vincent2011connection} shows that fitting the score model $\mathbf s_\theta$ to the score of the perturbation kernel $\nabla_{\mathbf x_t} \log p_{0t}(\mathbf x_t|\mathbf x_0, \mathbf y)$ is equivalent to implicit and explicit score matching \cite{hyvarinen2005estimation} under some regularity conditions. This technique is called \emph{denoising score matching} and essentially results in estimating 
\begin{align}
 \hspace*{-0.15em}\nabla_{\mathbf x_t}& \log p_{0t}(\mathbf x_t | \mathbf x_0, \mathbf y) = \nabla_{\mathbf x_t} \log \left[ |2\pi \sigma\mathbf I|^{-\frac{1}{2}} \mathrm{e}^{-\frac{\|\mathbf x_t - \boldsymbol \mu\|^2_2}{2\sigma^2}} \right] \\
&=\nabla_{\mathbf x_t} \log |2\pi \sigma(t)\mathbf I|^{-\frac{1}{2}} -  \nabla_{\mathbf x_t} \frac{\|\mathbf x_t - \boldsymbol\mu(\mathbf x_0,\mathbf y, t)\|^2_2}{2\sigma(t)^2} \\
\label{eq:before_subsitution}
&= -\frac{\mathbf x_t - \boldsymbol\mu(\mathbf x_0,\mathbf y, t)}{\sigma(t)^2}
\,,
\end{align}
where for simplicity we derived the score for the real and imaginary part of the complex normal distribution in Eq.~\eqref{eq:perturbation-kernel}, assuming they are independently distributed and each follows a real-valued multivariate normal distribution. Note that Eq.~\eqref{eq:before_subsitution} involves division by $\sigma(t)^2$, which has very small numerical values (including 0) around $t=0$. To avoid undefined values and numerical instabilities, we thus introduce a small minimum process time $t_\varepsilon$, as done previously in the literature~\cite{song2021sde}.

At each training step, the procedure can then be described as follows: 1) sample a random $t \sim \mathcal{U}[t_\varepsilon,T]$, 2) sample $(\mathbf x_0, \mathbf y)$ from the dataset, 3) sample $\mathbf z \sim \mathcal{N}_{\mathbb C}(\mathbf z; 0, \mathbf{I})$, and 4) sample $\mathbf x_t$ from Eq. \eqref{eq:perturbation-kernel} by effectively computing
\begin{align}
\label{eq:sample_gaussian}
\mathbf x_t =  \boldsymbol\mu(\mathbf x_0,\mathbf y, t) + \sigma(t) \mathbf z.
\end{align}
After passing $(\mathbf x_t, \mathbf y, t)$ to the score model, the final loss is an unweighted $L_2$ loss between the model output and the score of the perturbation kernel. By substituting Eq.~\eqref{eq:sample_gaussian} into Eq.~\eqref{eq:before_subsitution}, the overall training objective becomes,
\begin{equation}\label{eq:training-loss}
    \argmin_\theta \mathbb{E}_{t,(\mathbf x_0,\mathbf y), \mathbf z, \mathbf x_t|(\mathbf x_0,\mathbf y)} \left[
        \norm{\mathbf s_\theta(\mathbf x_t, \mathbf y, t) + \frac{\mathbf z}{\sigma(t)}}_2^2
    \right],
\end{equation}
where the expectation is approximated by sampling all random variables at each training step as described above. Note that due to the cancellation of $\boldsymbol\mu(\mathbf x_0,\mathbf y, t)$, the loss function does not explicitly involve $\mathbf y$, only as an input to the score model. This means that the score model is not tasked with estimating any portion of the environmental noise directly. Finally, the minimization is achieved by optimizing the parameters $\theta$ using stochastic gradient descent.

\subsection{Interpretation and limitations}
\label{sec:explorations}

\begin{figure*}
     \begin{subfigure}[t]{0.8\textwidth}
     \vskip 0pt
         \includegraphics[scale=0.71]{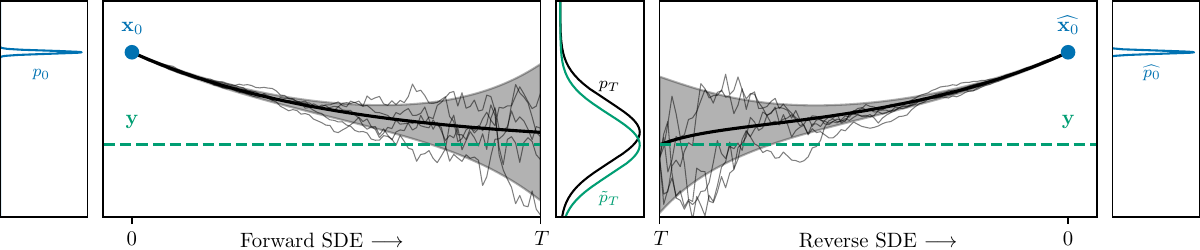}
     \end{subfigure}
     \hfill
     \begin{subfigure}[t]{0.17\textwidth}
    \vskip 0pt
         \includegraphics[scale=0.57]{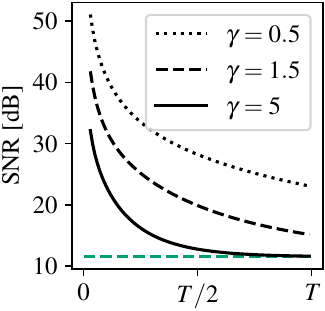}
     \end{subfigure}
     \caption{(Left) The forward and reverse process illustrated with a single scalar variable. The mean $\boldsymbol \mu$ (thick black line) of the forward process exponentially decays from clean speech $\mathbf x_0$ (blue) towards noisy speech $\mathbf y$ (green), and the standard deviation (shaded gray region) increases exponentially. The reverse process moves back to $\mathbf x_0$, starting from a slightly mismatched distribution $\tilde{p}_T$ which is centered around $\mathbf y$ rather than $\mathbf x_T$. Sample paths from both processes are shown as thin black lines. (Right) Time evolution of the \ac{snr} of the mean $\boldsymbol \mu$ (black) with respect to the \ac{snr} of $\mathbf y$ (green) for three different values of $\gamma$.}
     \label{fig:forward-process}
\end{figure*}

Let $p_t$ be the distribution of the perturbed data $\mathbf x_t$ from the diffusion process for a given dataset. Then its time evolution can be thought of as a continuum of distributions $\{p_t\}_{t\in[0,T]}$ which is determined by the drift and the diffusion coefficient of the forward \ac{sde}. 
For fixed $\mathbf x_0$ and $\mathbf y$, this time evolution can be described in close form using Eq.  \eqref{eq:perturbation-kernel}, which we illustrate in Fig.~\ref{fig:forward-process} for a one-dimensional case.
In the reverse process, the \ac{dnn} has the task of learning this continuous family of distributions starting from $\tilde {p}_T$ as defined in Eq. \eqref{eq:init_dist_reverse_process}. 
Due to the exponential increase of the diffusion coefficient in the forward process with initial condition $\sigma(0)^2 = 0$ the distribution $p_0$ essentially corresponds to the clean speech distribution, whereas the terminating distribution $p_T$ is a strongly corrupted version of the noisy speech distribution.
The particular characteristics in each noisy speech sample are strongly masked by the Gaussian white noise at $t=T$. Therefore, by learning the reverse process, the generative model learns a strong prior $p_0$ on clean speech, whereas the forward process terminates in a strongly corrupted distribution of the noisy speech, used as a weakly informative prior for generation. 
In Fig.~\ref{fig:forward-process}, we simulate five sample paths from the diffusion process. All sample paths of the forward process start exactly at $\mathbf x_0$ but exhibit starkly different trajectories at large $t$. The reverse process should then turn a high-variance sample $\mathbf  x_T$ back into a low-variance estimate of $\mathbf  x_0$.

Eq. \eqref{eq:mean} indicates that the mean $\boldsymbol \mu$ of the forward process exponentially decays from $\mathbf x_0$ to $\mathbf y$, which can also be seen in Fig. \ref{fig:forward-process} (thick black line). However, for finite $t$ it does not fully reach the corrupted speech $\mathbf y$ (dashed green line), particularly we have $\boldsymbol \mu(\mathbf x_0, \mathbf y, T) \neq \mathbf y$. Thus, the final distribution of the forward process $p_T$ exhibits a slight mismatch to the initial distribution of the reverse process $\tilde{p}_T$. We can make this mismatch arbitrarily small by either choosing a high stiffness parameter $\gamma$ or by increasing $\sigma_\text{max}$ to further smooth the density functions of both distributions.
However, increasing $\gamma$ would bring the mean close to $\mathbf y$ within a short time of the forward process, which may lead to an unstable reverse process because only the last steps are concerned with removing environmental noise. This effect can be seen in Fig. \ref{fig:forward-process} (right plot), where we plot the \ac{snr} of the process mean $\boldsymbol \mu$ averaged over 256 randomly selected files from the dataset for three values of $\gamma$. Note that we calculated the \ac{snr} in the time domain as the ratio of the power of clean speech to the power of environmental noise after inverting the non-linear amplitude transformation of Eq. \eqref{eq:spec-transform}. We see that while for $\gamma = 5$ the mismatch at $t=T$ becomes virtually zero, the change in \ac{snr} occurs mainly in the first half of the process. For $\gamma = 0.5$, on the other hand, the mismatch is already more than 10$\,\text{dB}$. However, the slope in the \ac{snr} is still apparent at the end of the process. Therefore, there is a trade-off to consider when choosing $\gamma$ which depends on the dataset to be used. Increasing $\sigma_\text{max}$ would come at the cost of more reverse iterations since the more white Gaussian noise is added, the less high-level information about the structure of the speech is preserved to serve as a guide in the reverse process. In the experiments, we choose a set of parameters based on empirical hyperparameter optimization.

\section{Numerical SDE Solvers}\label{sec:sampling}

There exist several computational methods to find numerical solutions for \acp{sde}, which are based on an approximation to discrete time steps. To this end, the interval $[0, T]$ is partitioned into $N$ equal subintervals of width $\Delta t=T/N$, which approximates the continuous formulation into the discrete reverse process $\{\mathbf x_T, \mathbf x_{T-\Delta t}, \dots, \mathbf x_0\}$. A common single-step method for solving this discretization is the Euler-Maruyama method. In each iteration step, the method refers to a previous state of the process and utilizes the drift and the Brownian motion to determine the current state.

In this work, we employ so-called predictor-corrector (PC) samplers proposed by Song et al. \cite{song2021sde}, which combine single-step methods for solving the reverse \ac{sde} with numerical optimization approaches such as annealed Langevin Dynamics \cite{song2019generative}. PC samplers consist of two parts, a predictor and a corrector. The predictor can be any single-step method that aims to solve the reverse process by iterating through the  reverse \ac{sde}. After each iteration step of the predictor, the current state of the process is refined by the corrector.
The correction is based on Markov chain Monte Carlo sampling and can be understood as a stochastic gradient ascent optimizer that adds at each iteration step a small amount of noise after taking a step in the direction of the estimated score.
One possible intuition about the use of stochastic correctors is that they allow the process state to escape local minima by the use of stochasticity. However, Karras et al.~\cite{karras2022elucidating} have recently argued that in the reverse process, stochasticity is only necessary to correct for numerical truncation errors of the predictor, a need that could be effectively circumvented by further improving the quality of the score model and predictor.

Another numerical way of approximating the reverse process is by solving the corresponding \emph{probability flow \ac{ode}},
\begin{equation}\label{eq:reverse-ode}
    \D{\mathbf x_t} =
        \left[
            \mathbf f(\mathbf x_t, \mathbf y) - \frac{1}{2}g(t)^2\mathbf s_\theta(\mathbf x_t, \mathbf y, t)
        \right] \D{t}\,,
\end{equation}
which is the associated \emph{deterministic process} of the stochastic reverse \ac{sde} in Eq. \eqref{eq:plug-in-reverse-sde}. It can be shown that for each diffusion process, there exists an \ac{ode} that describes the same marginal probability density $p_t(\mathbf x_t)$ \cite{song2021sde}. Enhancing the noisy or reverberant mixture is then based on solving this ODE. In Sec. \ref{sec:sampler_settings}, we also evaluate and compare this class of solvers for our task, specifically employing the Runge-Kutta method of fourth order with an error estimator of fifth order \cite{Dormand1980AFO}.

\section{Network Architecture}
\label{sec:architecture}

\begin{figure*}
     \centering
     \begin{subfigure}[b]{0.72\textwidth}
         \centering
         \includegraphics[scale=1]{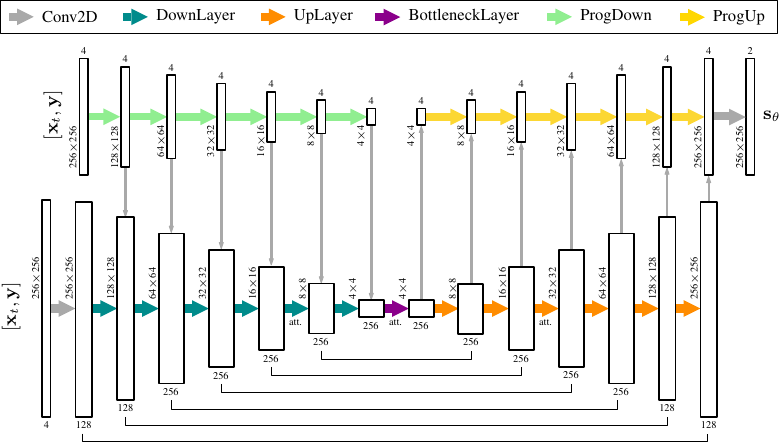}
         \caption{Feature maps at multiple resolutions.}
         \label{fig:feature_maps}
     \end{subfigure}
     \hfill
     \begin{subfigure}[b]{0.27\textwidth}
         \centering
         \includegraphics[scale=1]{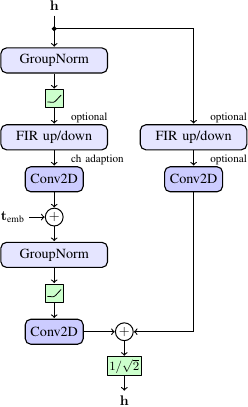}
    \caption{Residual block.}
    \label{fig:resnet-block}
     \end{subfigure}
     \caption{NCSN++ network architecture used as a score model $\mathbf s_\theta$: The architecture is based on a multi-resolution U-Net structure containing skip connections and an additional progressive growing path as shown in (a). Each up- and downsampling layer and the bottleneck layer consist of multiple residual blocks in series which are illustrated in (b).}
     \label{fig:ncsn++}
\end{figure*}

We utilize the \emph{Noise Conditional Score Network} (NCSN++) architecture \cite{song2021sde} for the score model $\mathbf s_\theta$ and adapt it for the use of complex spectrograms. For this purpose, we consider the real and imaginary parts of the complex input as separate channels, since the original network only works with real-valued numbers. Estimating both the real and imaginary parts of the score allows to generate complex spectrograms of clean speech. 

The network is based on a multi-resolution U-Net structure, which has been experimentally shown to be powerful for tasks such as generation and segmentation \cite{ronneberger2015unet}. In Fig. \ref{fig:feature_maps}, we illustrate the architecture by showing the feature maps at each resolution, indicating their spatial dimension and the corresponding number of channels. The transformations between the feature maps are represented by arrows, where the color of the arrow specifies the type of transformation (see the legend on top). We use Conv2D layers with a 3x3 kernel and stride 1 as input and output layers, and 1x1 Conv2D layers to aggregate information from the progressive growing path that we describe later. Up- and downsampling layers are based on residual network blocks which are taken from the BigGAN architecture \cite{brock2018large}, shown in Fig. \ref{fig:resnet-block}. A residual block consists of Conv2D layers with the same configuration as above, group normalization \cite{wu2018group}, up- or downsampling with finite impulse response (FIR) filters \cite{zhang2019making}, and the Swish activation function \cite{ramachandran2017swish}. Each upsampling layer consists of three residual blocks and each downsampling layer of two residual blocks in series with the last block performing the up- or downsampling. Global attention mechanisms \cite{vaswani2017attention} are added at a resolution of $16\times16$ and in the bottleneck layer to better learn global dependencies within the feature maps. 

To make the model time-dependent, information about the current progression of the diffusion process is fed into the network architecture. A common practice is to use Fourier-embeddings \cite{vaswani2017attention}, i.e., a learned projection that maps the scalar time coordinate $t$ to an $M$-dimensional vector $\mathbf t_\text{emb}$ that is integrated into every residual block as can be seen in Fig. \ref{fig:resnet-block}. 

In addition to the main feature extraction path of the multi-resolution U-Net structure, the network incorporates a so-called progressive growing of the input which is seen at the top of Fig. \ref{fig:feature_maps}. The idea is to provide a downsampled version of the input to every feature map in the contracting path, which has been successful in stabilizing high-resolution image generation \cite{karras2020analyzing}. Note that the downsampling operation in the progressive growing use shared weights for each resolution. The same procedure is also used in the expansive path where a progressive growing of the output is informed by the feature maps at each resolution, resulting in the final score estimate. 

\section{Experiments}

In this section, we describe the experimental setup for our speech enhancement and speech dereverberation experiments using the proposed method.

\subsection{Datasets}
\label{sec:datasets}

For the evaluation of the speech enhancement task we use two datasets, the WSJ0-CHiME3 dataset and the VB-DMD dataset, which are described below. The use of two datasets allows cross-dataset evaluation, i.e. the test is performed on the other dataset than the one used for training. 
This mismatched condition reveals information about how well the method generalizes to unseen data with different characteristics such as distinct noise types or different recording conditions. Moreover, to train and evaluate our proposed method on the dereverberation task, we create the WSJ0-REVERB dataset, which is also described below.

\paragraph{WSJ0-CHiME3} We create the WSJ0-CHiME3 data\-set using clean speech utterances from the Wall Street Journal (WSJ0) dataset \cite{datasetWSJ0}  and noise signals from the CHiME3 dataset \cite{barker2015third}. The mixture signal is created by randomly selecting a noise file and adding it to a clean utterance. Each utterance is used only once, and the \ac{snr} is  sampled uniformly between 0 and 20$\,$dB for the training, validation, and test set.

\paragraph{VB-DMD} We use the publicly available VoiceBank-DEMAND dataset (VB-DMD) \cite{valentini2016investigating} which is often used as a benchmark for single-channel speech enhancement. The utterances are artificially contaminated with eight real-recorded noise samples from the DEMAND database \cite{thiemann2013diverse} and two artificially generated noise samples (babble and speech shaped) at \acp{snr} of 0, 5, 10, and 15$\,$dB. The test utterances are mixed with different noise samples at \ac{snr} levels of 2.5, 7.5, 12.5, and 17.5$\,$dB. We split the training data into a training and validation set using speakers ``p226'' and ``p287'' for validation.

\paragraph{WSJ0-REVERB}

To create the WSJ0-REVERB data\-set, we use clean speech data from the WSJ0 dataset \cite{datasetWSJ0} and convolve each utterance with a simulated \ac{RIR}.  We use the \texttt{pyroomacoustics} engine \cite{Scheibler_2018} to simulate the \acp{RIR}. The reverberant room is modeled by sampling uniformly a $\T$ between 0.4 and 1.0 seconds.
A dry version of the room is generated with the same geometric parameters but a fixed absorption coefficient of 0.99, to generate the corresponding anechoic target. The resulting average \ac{DRR} is around -9$\,$dB.

\subsection{Instrumental evaluation metrics}
\label{sec:metrics}

To evaluate the performance of the proposed method we use standard metrics which we will describe in detail below. Metrics (a)-(d) employ full reference algorithms that rate the processed signal in relation to the clean reference signal using conventional digital signal analysis. On the other hand, metrics (e)-(g) are non-intrusive metrics that can be used to evaluate real recordings
when the clean reference is unavailable.

\paragraph{POLQA} The Perceptual Objective Listening Quality Analysis (POLQA) is an ITU-T standard that includes a perceptual model for predicting speech quality \cite{polqa2018}. The POLQA score takes values from 1 (poor) to 5 (excellent) as usual for \ac{mos}.

\paragraph{PESQ} The Perceptual Evaluation of Speech Quality (PESQ) is used for objective speech quality testing and is standardized in ITU-T P.862 \cite{rixPerceptualEvaluationSpeech2001}. Although it is the predecessor of POLQA, it is still widely used in the research community. The PESQ score lies between 1 (poor) and 4.5 (excellent) and there exist two variants, namely wideband PESQ and narrowband PESQ denoted as $\text{PESQ}_\text{nb}$. 

\paragraph{ESTOI} 
The Extended Short-Time Objective Intelligibility (ESTOI) is an instrumental measure for predicting the intelligibility of speech subjected to various kinds of degradation \cite{jensen2016algorithm}. The metric is normalized and lies between 0 and 1, with higher values indicating better intelligibility.

\paragraph{SI-SDR, SI-SIR, SI-SAR} 
Scale-Invariant (SI-) Signal-to-Distortion Ratio (SDR), Signal-to-Interference Ratio (SIR), and Signal-to-Artifact Ratio (SAR) are standard evaluation metrics for single-channel speech enhancement and speech separation\cite{leroux2018sdr}. They are all measured in dB, with higher values indicating better performance.

\paragraph{DNSMOS} The Deep Noise Suppression \ac{mos} (DNSMOS) is a reference-free metric to evaluate perceptual speech quality \cite{reddy2021dnsmos}. The evaluation method uses a \ac{dnn} that is trained on human ratings obtained by using an online framework for listening experiments \cite{naderi2020open} based on ITU-T P.808 \cite{itu-t-rec808}.

\paragraph{SIG, BAK, OVRL} The non-intrusive speech quality assessment model DNSMOS P.835 \cite{reddy2022dnsmos} is based on a listening experiment according to ITU-T P.835 \cite{itu-t-rec835} and provides three \ac{mos} scores: speech quality (SIG), background noise quality (BAK), and the overall quality (OVRL) of the audio. 

\paragraph{WVMOS} Wav-to-Vec \ac{mos} (WVMOS) \cite{andreev2022hifi} is a MOS prediction method for speech quality evaluation using a fine-tuned wav2vec2.0 model \cite{baevski2020wav2vec}.

\subsection{Listening Experiment}
Instrumental evaluation metrics do not always correlate to human perception because there are many aspects of perception that are very difficult to capture by computational means. Therefore, we conduct a MUSHRA listening experiment \cite{mushra2014method} with ten participants using the webMUSHRA framework \cite{schoeffler2018webmushra}. The participants were asked to rate the overall quality of twelve randomly sampled examples from the WSJ0-CHiME3 test set as reconstructed by the compared algorithms. The results are reported on a quality scale from 0 to 100.

\subsection{Hyperparameters and training configuration}

\paragraph{Input representation}
We convert each audio input with sampling rate $16\,\text{kHz}$ into a complex-valued \ac{stft} representation using a window size of 510, resulting in $F=256$, a hop length of 128 (i.e. approximately 75\% overlap), and a periodic Hann window. To process multiple examples for batch training, the length of each spectrogram is trimmed to $K=256$ \ac{stft} time frames, with start and end times selected randomly at each training step. For the spectrogram transformation in Eq. \eqref{eq:spec-transform}, we have chosen $\alpha = 0.5$ and $\beta = 0.15$ empirically. 

\paragraph{Stochastic process}
The \ac{sde} in Eq.~\eqref{eq:ouve-sde} is parameterized with $\sigma\submin = 0.05, \sigma\submax = 0.5$, and $\gamma = 1.5$ based on hyperparameter optimization with grid search. 

\paragraph{Training configuration}

We train the \ac{dnn} on four Quadro RTX 6000 (24 GB memory each) for 160 epochs using the distributed data-parallel (DDP) approach in PyTorch Lightning \cite{falcon2019pytorch}, which takes about one day. We use the Adam optimizer~\cite{kingma2015adam} with a learning rate of $10^{-4}$ and an effective batch size of $4 \times 8 = 32$. We track an exponential moving average of the DNN weights with a decay of 0.999, to be used for sampling~\cite{song2020improved}. We log the average PESQ value of 20 randomly chosen examples from the validation set during training and select the best-performing model for evaluation.

\subsection{Sampler settings}
\label{sec:sampler_settings}

To find optimal sampler settings for the reverse process, we run a hyperparameter search using the VB-DMD dataset.

\paragraph{Sampler type} 
We investigate which choice of sampler yields the best speech enhancement performance, comparing the PC sampler with different numbers of corrector steps and an ordinary ODE sampler as described in Sec. \ref{sec:sampling}. In Tab. \ref{tab:sampler_type} we can see that use of one correction step in the PC sampler seems to be advantageous, but the use of two steps does not lead to a further increase in performance. Thus, we decide to use the PC sampler with one corrector step for the evaluation. However, it should be noted that the use of one correction step doubles the \ac{nfe} of the sampler. The \emph{function} being the expensive score model, this results in an average \ac{rtf} of 1.77, i.e., 1$\,$sec of audio requires 1.77$\,$sec of processing\footnote{Average processing time for 10 audio files on an NVIDIA GeForce RTX 2080 Ti GPU, in a machine with an Intel Core i7-7800X CPU @ 3.50GHz.\label{footnote_rtf}}. Comparing the PC sampler with the ODE sampler, we find that the PC sampler performs better in both metrics. However, with suitable settings, the ODE sampler requires only 14 \ac{nfe} on average which results in an improved \ac{rtf} of only 0.46.

\begin{table}[t]
    \centering
    \caption{Results for different sampler configurations tested on VB-DMD with the average number of function evaluations (NFE) and the respective average real-time factor (RTF)\footref{footnote_rtf}.}
    \begin{tabular}{cccc|cc}
        \toprule 
        Type  & Sampler settings & NFE & RTF & PESQ  & SI-SDR$\,$[dB] \\
        \midrule
        PC  & 0 corrector steps & 30 & 0.89 & 2.80 & 15.38 \\
        PC  & 1 corrector steps & 60 & 1.77 & 2.93 & 17.35 \\
        PC  & 2 corrector steps & 90 & 2.65 & 2.92 & 17.52 \\
        \midrule
        ODE & atol=$10^{-1}$, rtol=$10^{-1}$ & 14 & 0.46 & 2.78 & 12.83  \\
        ODE & atol=$10^{-6}$, rtol=$10^{-3}$  & 49 & 1.55 & 2.71 & 12.76 \\
        \bottomrule
    \end{tabular}
    \label{tab:sampler_type}
\end{table}

\begin{figure}
     \centering
     \begin{subfigure}[b]{0.49\columnwidth}
     \centering
         \includegraphics[scale=0.7]{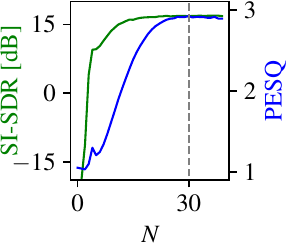}
        \caption{Varying $N$, fix $r=0.33$}
        \label{fig:reverse_steps}
     \end{subfigure}
     \begin{subfigure}[b]{0.49\columnwidth}
      \centering
         \includegraphics[scale=0.7]{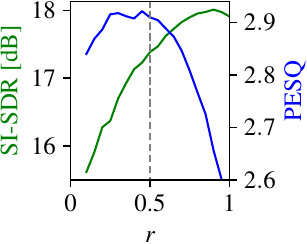}
        \caption{Varying $r$, fix $N=30$}
        \label{fig:step_size_corrector}
     \end{subfigure}
        \caption{Model performance in PESQ and SI-SDR as a function of (a) the number of reverse steps $N$ and (b) the step size parameter $r$ in the annealed Langevin corrector.}
        \label{fig:fig:sampler_hyperparameters}
\end{figure}

\paragraph{Number of reverse steps N}
The number of reverse steps $N$ can be used to set a balance between the computational effort and the performance of the model. In Fig. \ref{fig:reverse_steps}, we show the speech enhancement performance as a function of $N$. It can be seen that SI-SDR starts to stagnate earlier than PESQ. We opt for a value of $N=30$, at which both metrics show no further increase in performance.

\paragraph{Step size in corrector}
In Fig. \ref{fig:step_size_corrector}, we vary the step size $r$ of the annealed Langevin dynamics in the corrector. Interestingly, this parameter represents a compromise between PESQ and SI-SDR. We choose $r=0.5$ to achieve a maximum PESQ value while still obtaining a good value for SI-SDR.

\subsection{Baselines}
We compare the performance of our proposed method with four generative and four discriminative baselines which we describe in more detail below. All methods are re-trained by us, except for DVAE, MetricGAN+, and CDiffuSE on VB-DMD, for which we obtained the pre-trained model from the authors who used the exact same training data.  

\paragraph{STCN \cite{richter2020speech}}
A generative \ac{vae}-based speech enhancement method which uses a stochastic temporal convolutional network (STCN) \cite{aksan2018stcn} that allows the latent variables to have both hierarchical and temporal dependencies. The parameters of the noise model and the latent variables are estimated using a Monte Carlo expectation maximization (MCEM) algorithm.

\paragraph{DVAE \cite{bie2022unsupervised}}
Generative speech enhancement method based on an  unsupervised dynamical \ac{vae} (DVAE) \cite{girin2021dynamical} which models temporal dependencies between successive observable and latent variables. Parameters are updated at test time using a variational expectation maximization (VEM) method where the encoder is fine-tuned using stochastic gradient ascent. 

\paragraph{CDiffuSE \cite{lu2022conditional}} Most related to our proposed method is CDiffuSE, a generative speech enhancement method based on a conditional diffusion process defined in the time domain. 

\paragraph{SGMSE \cite{welker2022speech}} 
Score-based Generative Model for Speech Enhancement (SGMSE) is our previous publication on which the proposed method is based. The main difference is that it uses a deep complex U-Net \cite{choi2019phase} instead of the NCSN++ architecture as the score model.

\paragraph{MetricGAN+ \cite{fu2021metricgan+}} 
A discriminative speech enhancement method that uses a generator network for mask-based prediction of clean speech and introduces a discriminator network trained to approximate the PESQ score.

\paragraph{Conv-TasNet \cite{luo2019conv}}
An end-to-end neural network that estimates a mask that is used for filtering a learned representation of the noisy mixture. The filtered representation is transformed back to the time domain by a learned decoder.

\paragraph{GaGNet \cite{li2022gagnet}}
This neural network is trained on a hybrid complex-domain and magnitude-domain regression objective for single-channel dereverberation. It uses so-called \enquote{glance} and \enquote{gaze} (GaG) modules, which respectively perform a coarse estimation of the magnitude and refine it with phase estimation in the complex domain.

\paragraph{TCN+SA+S \cite{zhao2020speechdereverberationwithtcnandselfattention}}
This single-channel dereverberation approach uses a self-attention module to extract features from the input magnitude. This representation is then used by a temporal convolutional network followed by a single-layer convolutional smoother that outputs a magnitude estimate, which is used as the training objective. Griffin-Lim iterations are used to reconstruct the phase.

\section{Results}

\subsection{Speech Enhancement}
\label{sec:results:se}

In Tab.~\ref{tab:results:wsj0}, we report the speech enhancement results on the WSJ0-CHiME3 test set for the matched and mismatched condition, i.e. when the training set was also WSJ0-CHiME3 or when the training set was VB-DMD. We compare our proposed method, which we call \emph{SGMSE+}, with selected baseline methods and sort the results by the type of algorithm, which is either generative or discriminative. Considering the matched condition in the upper half of Tab.~\ref{tab:results:wsj0}, we see that SGMSE+ outperforms all other generative methods in all metrics.
Note that STCN and RVAE are both unsupervised speech enhancement methods, i.e. they are trained on clean speech only (WSJ0 or VB). RVAE shows competitive results for SI-SAR, however, its VEM optimization algorithm is very time-consuming due to the fine-tuning of the encoder at test time, resulting in a \ac{rtf} of~$>$10000. This is significant in contrast to STCN with a \ac{rtf} of 0.64 and SGMSE+ with a \ac{rtf} of 1.77\footref{footnote_rtf}. Although both VAE-based methods model temporal dependencies, they are limited in their ability to produce high-quality speech, likely due to the dimensionality reduction of the latent variable and the encoder's sensitivity to noisy input, which causes the latent variable to be incorrectly initialized \cite{fang2021variational}.

\begin{table*}[t]
    \centering
    \caption{Speech enhancement results obtained for WSJ0-CHiME3 under matched and mismatched training conditions. Values indicate mean and standard deviation. Methods are sorted by the algorithm type, generative (G) or discriminative (D).}
\resizebox{\textwidth}{!}{

\begin{tabular}{l|cc|cccccc|c}
\toprule
Method & Type & Training set & POLQA & PESQ & ESTOI & SI-SDR [dB] & SI-SIR [dB] & SI-SAR[dB] & DNSMOS \\
\midrule
Mixture & - & - & $2.63 \pm 0.67$ & $1.70 \pm 0.49$ & $0.78 \pm 0.14$ & $10.0 \pm 5.7$ & $10.0 \pm 5.7$ & - & $3.11 \pm 0.39$ \\
\midrule
STCN \cite{richter2020speech} & G & WSJ0 & $2.64 \pm 0.68$ & $2.01 \pm 0.55$ & $0.81 \pm 0.12$ & $13.5 \pm 4.7$ & $18.7 \pm 5.5$ & $15.4 \pm 4.7$ & $3.34 \pm 0.37$ \\
RVAE \cite{bie2022unsupervised} & G & WSJ0 & $2.97 \pm 0.63$ & $2.31 \pm 0.55$ & $0.85 \pm 0.11$ & $15.8 \pm 5.0$ & $21.6 \pm 6.1$ & $17.6 \pm 4.9$ & $3.61 \pm 0.29$ \\
CDiffuse \cite{lu2022conditional} & G & WSJ0-C3 & $3.08 \pm 0.58$ & $2.27 \pm 0.51$ & $0.83 \pm 0.09$ & $\:\:9.2 \pm 2.3$ & $19.8 \pm 5.9$ & $10.0 \pm 2.3$ & $3.43 \pm 0.32$ \\
SGMSE \cite{welker2022speech} & G & WSJ0-C3 & $2.98 \pm 0.60$ & $2.28 \pm 0.57$ & $0.86 \pm 0.09$ & $14.8 \pm 4.3$ & $25.4 \pm 5.6$ & $15.3 \pm 4.2$ & $3.70 \pm 0.27$ \\
SGMSE+ & G & WSJ0-C3 & $\mathbf{3.73 \pm 0.53}$ & $2.96 \pm 0.55$ & $0.92 \pm 0.06$ & $18.3 \pm 4.4$ & $\mathbf{31.1 \pm 4.6}$ & $18.6 \pm 4.5$ & $\mathbf{3.99 \pm 0.19}$ \\
\midrule
MetricGAN+ \cite{fu2021metricgan+} & D & WSJ0-C3 & $3.52 \pm 0.61$ & $\mathbf{3.03 \pm 0.45}$ & $0.88 \pm 0.08$ & $10.5 \pm 4.5$ & $24.5 \pm 5.1$ & $10.7 \pm 4.6$ & $3.67 \pm 0.30$ \\
Conv-TasNet \cite{luo2019conv} & D & WSJ0-C3 & $3.65 \pm 0.54$ & $2.99 \pm 0.58$ & $\mathbf{0.93 \pm 0.05}$ & $\mathbf{19.9 \pm 4.3}$ & $29.2 \pm 4.6$ & $\mathbf{20.6 \pm 4.5}$ & $3.79 \pm 0.27$ \\
\midrule
STCN \cite{richter2020speech} & G & VB & $2.53 \pm 0.66$ & $1.80 \pm 0.45$ & $0.79 \pm 0.12$ & $11.9 \pm 4.5$ & $17.3 \pm 4.9$ & $13.8 \pm 4.6$ & $3.40 \pm 0.34$ \\
RVAE \cite{bie2022unsupervised} & G & VB & $2.84 \pm 0.61$ & $2.08 \pm 0.49$ & $0.82 \pm 0.11$ & $13.9 \pm 4.8$ & $19.5 \pm 5.9$ & $15.8 \pm 4.7$ & $3.52 \pm 0.31$ \\
CDiffuse \cite{lu2022conditional} & G & VB-DMD & $2.15 \pm 0.57$ & $1.79 \pm 0.42$ & $0.71 \pm 0.11$ & $\:\:3.2 \pm 3.2$ & $21.8 \pm 7.0$ & $\:\:3.4 \pm 3.2$ & $3.17 \pm 0.29$ \\
SGMSE \cite{welker2022speech} & G & VB-DMD & $2.66 \pm 0.58$ & $1.94 \pm 0.47$ & $0.81 \pm 0.11$ & $13.3 \pm 4.3$ & $23.5 \pm 6.0$ & $13.8 \pm 4.2$ & $3.76 \pm 0.25$ \\
SGMSE+ & G & VB-DMD & $\mathbf{3.43 \pm 0.61}$ & $\mathbf{2.48 \pm 0.58}$ & $\mathbf{0.90 \pm 0.07}$ & $\mathbf{16.2 \pm 4.1}$ & $\mathbf{28.9 \pm 4.6}$ & $\mathbf{16.4 \pm 4.1}$ & $\mathbf{4.00 \pm 0.19}$ \\
\midrule
MetricGAN+ \cite{fu2021metricgan+} & D & VB-DMD & $2.47 \pm 0.67$ & $2.13 \pm 0.53$ & $0.76 \pm 0.12$ & $\:\:6.8 \pm 3.1$ & $22.9 \pm 4.9$ & $\:\:7.0 \pm 3.1$ & $3.51 \pm 0.29$ \\
Conv-TasNet \cite{luo2019conv} & D & VB-DMD & $3.13 \pm 0.60$ & $2.40 \pm 0.53$ & $0.88 \pm 0.08$ & $15.2 \pm 3.9$ & $26.5 \pm 4.6$ & $15.6 \pm 4.0$ & $3.68 \pm 0.30$ \\
\bottomrule
\end{tabular}

}
\label{tab:results:wsj0}
\end{table*}

Comparing SGMSE+ to our previous model SGMSE, we find a significant improvement, especially for the perceptual metrics. We report improvements of 0.75 for POLQA and 0.68 for PESQ. This shows that the proposed generative diffusion process benefits significantly from the adapted network architecture. In our previous paper \cite{welker2022speech}, we have already shown improvements over CDiffuSE for SGMSE in SI-SDR and SI-SAR, which we now back up with also reporting an improvement in ESTOI and DNSMOS and on par results in PESQ and POLQA. With SGMSE+, these improvements become even more significant, e.g. with 0.65 improvement in POLQA and 9.1 dB in SI-SDR compared to CDiffuSE. In qualitative analysis, we found that SGMSE+ is more accurate than CDiffuSE in preserving the high frequencies of fricatives after the completion of the reverse process. 
To compensate for that, CDiffuSE combines the enhanced files with the original noisy speech signal at a ratio of 0.2 for the final prediction \cite{lu2022conditional}. This results in a trade-off between noise removal and the conservation of the signal. In our proposed approach, on the other hand, we found no significant suppression of high frequencies after completing the reverse process. Therefore, it is not necessary to mix back the noisy mixture to improve the signal quality, resulting in a significantly higher SI-SIR.

The comparison with Conv-TasNet and MetricGAN+ shows that SGMSE+ can keep up with the performance of discriminative methods and even surpasses them in terms of POLQA, SI-SIR, and DNSMOS.  
Discriminative methods are based on regression problems that optimize certain point-wise loss functions between the corrupted speech and a clean speech reference. For Conv-TasNet and MetricGAN+ these loss functions correspond to established intrusive metrics, namely SI-SDR for Conv-TasNet and PESQ for MetricGAN+. Note that both these discriminative methods shine in particular on the respective metric they used as a loss function.
In contrast, generative methods like SGMSE+ are usually not trained to achieve the exact reconstruction of the reference clean speech but rather aim at generating a realization of speech that is on the manifold of clean speech. 
Thus, we suggest the use of non-intrusive metrics as a complementary measure since they allow an estimation of speech quality without relying on the exact reconstruction of a reference signal. 
In fact, for the non-intrusive metric DNSMOS our proposed method yields a significantly higher value than the discriminative baselines, indicating the strong ability of our generative model to generate high-quality clean speech.

Looking at the results for the mismatched condition in the bottom half of Tab.~\ref{tab:results:wsj0}, a general trend of decreasing metrics can be seen for all methods when compared to the corresponding values of the matched condition. This was to be expected since particular properties of the mismatched test set, such as distinct  noise types or different recording characteristics of the clean speech have not been seen during training. However, generative methods generally show less degradation in the mismatched condition than discriminative methods. CDiffuSE is an exception, as this method shows significant degradation in the mismatched case. Informal listening reveals a problem with gain control, which is evident in strong volume fluctuations in the enhanced files. Furthermore, we see that SGMSE+ outperforms all other methods in all metrics under this condition, which shows the ability of our proposed method to generalize well. 

\begin{figure}[t]
    \centering
 \includegraphics[scale=0.7]{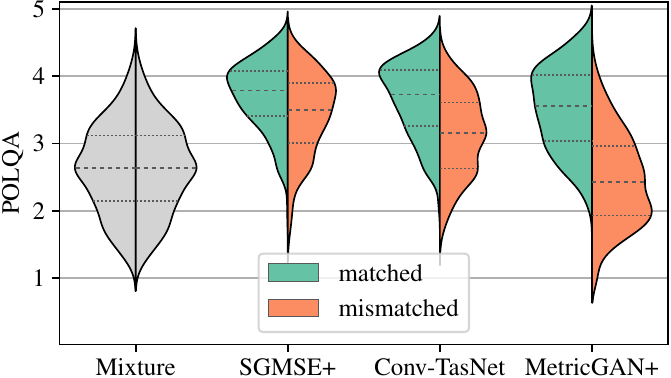}
    \caption{Violin plots showing POLQA results for the matched and the mismatched condition with dashed and dotted lines representing median and quartiles, respectively.}
    \label{fig:violin_plot}
\end{figure}

Complementary to the average results above, we present in Fig. \ref{fig:violin_plot} violin plots of the full distribution of the POLQA scores obtained for SGMSE+, Conv-TasNet, MetricGAN+, and the noisy mixture for reference. For each method, the distributions are plotted side by side for the matched and mismatched conditions, so that the ability to generalize can be inferred from the horizontal alignment between both distributions. It can be seen that both distributions for SGMSE+ are relatively similar, whereas they are skewed for Conv-TasNet and especially for MetricGAN+.

\begin{figure}[t]
    \centering
 \includegraphics[scale=0.7]{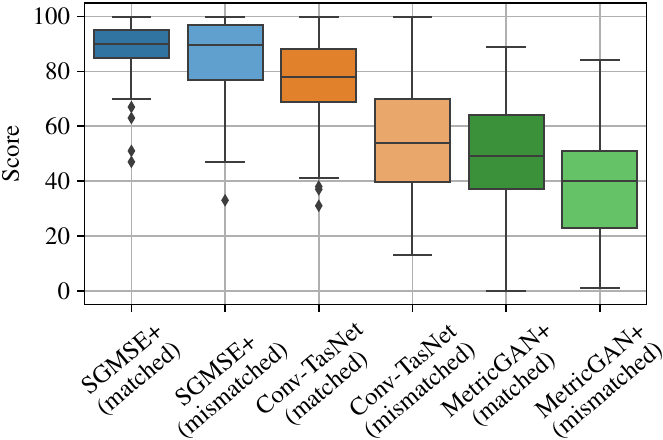}
    \caption{Boxplot showing the results of the MUSHRA listening experiment with ten participants on twelve randomly selected examples.}
    \label{fig:listening_experiment}
\end{figure}

In Fig. \ref{fig:listening_experiment}, we report the results of the MUSHRA listening experiment in a boxplot. On average, the ten participants rated the overall quality of our proposed approach with the highest score. In addition, our method remains fairly robust when the model was trained on a different training set, while discriminative methods show much stronger degradation for the mismatched condition. This also corresponds with the results of the non-intrusive metric DNSMOS in Tab. \ref{tab:results:wsj0} and thus supports the use of non-intrusive methods for instrumental evaluation. Interestingly, MetricGAN+ was only rated with a median score less than 50 for the matched condition, although the method performed best among all baselines for PESQ (see Tab. \ref{tab:results:wsj0}). This reveals the discrepancy between the use of instrumental metrics for evaluation and people's actual perceptions. We suspect that MetricGAN+ has simply learned to utilize the internal operations of the PESQ algorithm to obtain a high value in this metric, neglecting the naturalness of the clean speech estimate. In fact, listening to the enhanced files, it can be recognized that the energy of the speech signal estimated by MetricGAN+ is mainly concentrated in the low- and mid-frequency area of the spectrogram, while high frequencies are strongly attenuated. 

Listening to the enhanced files of our method, we notice that at very low input \acp{snr}, some \enquote{vocalizing} artifacts with very poor articulation and no linguistic meaning are occasionally produced. In other examples, we find that breathing sounds or speech-like sounds were generated in noisy regions where no speech was originally present. These artifacts may also explain the outliers of our method in the listening experiment (see Fig.~\ref{fig:listening_experiment}). For the matched condition, for example, the two lowest outliers come from the same utterance with clearly noticeable vocalizing artifacts. We hypothesize that these artifacts can be linked to the generative nature of the proposed approach. Indeed, for very noisy inputs, the score model may erroneously identify noise energy in some \ac{tf} areas as corrupted speech. The reverse diffusion process then produces speech where it did not originally exist. We argue that this behavior could be mitigated if some conditioning with respect to speech activity and phoneme identity would be added to the score model.

Finally, Tab. \ref{tab:results:vb} lists the results for the standardized VB-DMD dataset. This has the advantage that one can take values from other methods and copy them from the corresponding papers for a quick algorithmic comparison. It can be seen that SGMSE+ outperforms all other generative baselines, further narrowing the performance gap with discriminative methods that currently lead the benchmark based on PESQ, including recent approaches such as \cite{yu2022dbt} and \cite{cao22_interspeech}. 
It should however be noted that PESQ formally requires a minimum file length of 3.2$\,$sec according to P.862.3~\cite{PESQguide}, which is not the case for most files in VB-DMD~\cite{valentini2016investigating}.

\begin{table}[t]
    \centering
    \caption{Speech enhancement results obtained for VB-DMD. Models marked with an asterisk ($^*$) are additional baselines with values taken from the corresponding papers.}
    \resizebox{\columnwidth}{!}{
    
\begin{tabular}{l|cccc|c}
\toprule
Method & PESQ & $\text{PESQ}_\text{nb}$ & ESTOI & SI-SDR & DNSMOS \\
\midrule
Mixture & $1.97$ & $2.88$ & $0.79$ & $\:\:8.4$ & $3.09$ \\
\midrule
SEGAN$^*$ \cite{pascual2017segan} & 2.16 & - & - & - & - \\
RVAE \cite{bie2022unsupervised} & $2.43$ & $3.11$ & $0.81$ & $16.4$ & $3.30$ \\
MetricGAN-U$^*$ \cite{fu2022metricgan} & 2.45 & - & 0.77 & $\;\,$8.2 & - \\
CDiffuse \cite{lu2022conditional} & $2.52$ & $3.31$ & $0.79$ & $12.4$ & $3.09$ \\
SGMSE \cite{welker2022speech} & $2.28$ & $3.22$ & $0.80$ & $16.2$ & $3.46$ \\
SGMSE+ & $2.93$ & $\mathbf{3.66}$ & $\mathbf{0.87}$ & $17.3$ & $\mathbf{3.56}$ \\
\midrule
UMX$^*$ \cite{uhlich2020open} & 2.35 & - & 0.83 & 14.0 & - \\
Conv-TasNet \cite{luo2019conv} & $2.63$ & $3.42$ & $0.85$ & $\mathbf{19.1}$ & $3.37$ \\
MetricGAN+ \cite{fu2021metricgan+} & $\mathbf{3.13}$ & $3.63$ & $0.83$ & $\:\:8.5$ & $3.37$ \\
\bottomrule
\end{tabular}

}
\label{tab:results:vb}
\end{table}

Investigating whether phase estimation has actually been improved with the modeling of the complex coefficients, we use the noisy phase in place of the estimated phase which does not show a significant performance difference. This is also in line with a recent study on the role of phase enhancement, where it has been shown that the impact of phase enhancement is rather small for $\sim$32$\,$ms spectral analysis frames but increasingly large with shorter frame lengths \cite{peer2022phase}. 

\subsection{Dereverberation}
\label{sec:results:derev}

We report in Tab.~\ref{tab:results:dereverberation} the performance of our approach when trained and tested on a single-channel dereverberation task. We compare with SGMSE \cite{welker2022speech} and three discriminative baselines, namely Conv-TasNet \cite{luo2019conv}, GaGNet \cite{li2022gagnet} and TCN+SA+S \cite{zhao2020speechdereverberationwithtcnandselfattention}.

\begin{table*}[t]
    \centering
    \caption{Single-channel dereverberation results obtained for WSJ0-REVERB test set. Values indicate mean and standard deviation. Methods are sorted by the algorithm type which is either generative (G) or discriminative (D).}
    \begin{tabular}{l|c|cccccc}
        \toprule 
        Method & Type & POLQA & PESQ & ESTOI & SI-SDR & SI-SIR & SI-SAR \\
        \midrule
    Mixture & - & 1.76 $\pm$ 0.29 & 1.36 $\pm$ 0.19 & 0.46 $\pm$ 0.12 & -7.3 $\pm$ 5.5 & -7.5 $\pm$ 5.4 & - \\
     \midrule
SGMSE \cite{welker2022speech} & G & 1.79 $\pm$ 0.28 & 1.35 $\pm$ 0.15 & 0.57 $\pm$ 0.07 & -7.4 $\pm$ 5.8 & -1.1 $\pm$ 7.0 & -6.2 $\pm$ 5.5$\,$ \\
SGMSE+ & G & \textbf{3.24 $\pm$ 0.46} & \textbf{2.66 $\pm$ 0.48} &\textbf{0.84 $\pm$ 0.07} &\textbf{ 1.6 $\pm$ 7.8 }& $\;\,$ 9.4 $\pm$ 10.2 & \textbf{2.3 $\pm$ 7.2} \\
 \midrule
Conv-TasNet \cite{luo2019conv} & D & 2.41 $\pm$ 0.52 & 1.84 $\pm$ 0.42 & 0.73 $\pm$ 0.10 & $\;$\textbf{1.6 $\pm$ 5.3} & \textbf{12.1 $\pm$ 5.1} & 1.9 $\pm$ 5.4  \\
TCN+SA+S \cite{zhao2020speechdereverberationwithtcnandselfattention} & D & 2.92 $\pm$ 0.33 & 2.29 $\pm$ 0.36 & 0.79 $\pm$ 0.05 & -4.4 $\pm$ 5.3 & -2.3 $\pm$ 5.2 & -0.6 $\pm$ 5.1 $\,$ \\
GaGNet \cite{li2022gagnet} & D & 2.62 $\pm$ 0.47 & 1.98 $\pm$ 0.46 & 0.73 $\pm$ 0.08 & -0.6 $\pm$ 4.9 & $\,$ 6.1 $\pm$ 3.9 & 0.4 $\pm$ 5.1  \\
    \bottomrule
    \end{tabular}
    \label{tab:results:dereverberation}
\end{table*}

Our proposed SGMSE+ approach performs particularly well in terms of instrumental metrics compared to all other baseline models. The low average input \ac{DRR} of -9$\,$dB constitutes a real challenge for discriminative approaches, which do not manage to separate the reverberation from the target without distorting the target signal, resulting in low-quality scores. On the other hand, our approach benefits from generative modeling and is able to reconstruct speech with very high quality in most cases. When comparing our previous SGMSE model~\cite{welker2022speech} with SGMSE+, we see that for speech dereverberation, the method benefits greatly from the improved network architecture. This effect is even more significant than for additive background noise removal in the speech enhancement task.

In particular, using the proposed approach SGMSE+ on a single-channel dereverberation task does not produce any of the vocalized artifacts observed in the speech enhancement experiments for low input \acp{snr}. Although the reverberant signal is formally decorrelated in the time domain from the target by the randomness of reflections across the room, it still originates from the dry speech source. Therefore, we conjecture that the score model effectively detects whether the energy in a particular time-frequency area is associated with the clean speech nearby that needs to be reconstructed.

\subsection{Evaluation on real data}

Complementing the experiments using simulated data, we evaluate the speech enhancement performance on real-world noisy recordings. For real-world noisy recordings, there exists no clean speech reference. Thus, we can only non-intrusive metrics to evaluate the perceptual speech quality which we describe in Sec. \ref{sec:metrics} (e)-(g). For the evaluation, we use 300 files from the test set of the Deep Noise Suppression (DNS) Challenge 2020 \cite{reddy2020interspeech}. In Tab. \ref{tab:results:dns}, we report the results for models that were trained on VB-DMD. It turns out that our proposed method performs better than all other methods in all non-intrusive metrics, demonstrating its robustness to real-world noisy examples. Interestingly, a trend of degradation in speech quality (SIG) can be observed for the discriminative methods, whereas all generative models improve this metric with respect to the mixture. For the background noise quality (BAK) metric, on the other hand, discriminative models seem to perform well, yet our proposed method performs superior. It is important to note that non-intrusive metrics do not require a corresponding clean reference signal and only assess speech quality based on the method's estimate. We hypothesize that our generative model works well on these metrics, as it was trained to generate clean speech. However, ``vocalizing'' artifacts as mentioned above or phonetic confusions may not be captured with these metrics.

We provide on our project page\footnote{\url{https://uhh.de/inf-sp-sgmse}} some listening examples for all evaluated tasks. Furthermore, we include real reverberant examples from the MC-WSJ-AV dataset \cite{lincoln2005multi}.

\begin{table}[t]
    \centering
    \caption{Speech enhancement results obtained for real-world noisy recordings from the DNS Challenge 2020 test set.}
    \resizebox{\columnwidth}{!}{

\begin{tabular}{l|ccccc}
\toprule
Method & DNSMOS & SIG & BAK & OVRL & WVMOS \\
\midrule
Mixture & $3.05$ & $3.05$ & $2.51$ & $2.26$ & $1.12$ \\
\midrule
RVAE \cite{bie2022unsupervised} & $3.29$ & $3.16$ & $2.91$ & $2.44$ & $1.87$ \\
CDiffuse \cite{lu2022conditional} & $3.14$ & $3.15$ & $3.19$ & $2.55$ & $1.86$ \\
SGMSE \cite{welker2022speech} & $3.38$ & $3.22$ & $3.02$ & $2.52$ & $1.80$ \\
SGMSE+ & $\mathbf{3.64}$ & $\mathbf{3.42}$ & $\mathbf{3.82}$ & $\mathbf{3.04}$ & $\mathbf{2.54}$ \\
\midrule
Conv-TasNet \cite{luo2019conv} & $3.07$ & $2.87$ & $3.59$ & $2.52$ & $2.07$ \\
MetricGAN+ \cite{fu2021metricgan+} & $3.26$ & $2.88$ & $3.39$ & $2.45$ & $1.52$ \\
\bottomrule
\end{tabular}

}
\label{tab:results:dns}
\end{table}

\section{Conclusions}

In this work, we built upon our existing work \cite{welker2022speech} that uses a novel stochastic diffusion process to design a generative model for speech enhancement in the complex \ac{stft} domain. We presented an extended theoretical analysis of the underlying score-based generative model and derived in detail the objective function used for training. In further explorations, we considered the time evolution of the conditional diffusion process which revealed a slight mismatch between the forward and reverse process, which can be adjusted with a careful parameterization of the forward \ac{sde}. 

By using an adopted network architecture, we were able to significantly improve the performance compared to our previous model. In addition, we trained and evaluated the proposed method on the task of speech dereverberation and show significantly superior performance compared to discriminative baseline methods. Hence, we showed that with our proposed method, a single framework can be used to train individual models for different distortion types. 
For the task of speech enhancement, we evaluated performance under matched and mismatched conditions, i.e. when the training and test data were taken from the same or different corpora. For the matched condition, the proposed generative speech enhancement method performs on par with competetive discriminative methods. For the mismatched condition, our method shows strong generalization capabilities and outperforms all baselines in all metrics, as confirmed by a listening experiment. In very adverse conditions, however, we observe that the proposed method sometimes introduces vocalizing and breathing artifacts. We argue that these could be mitigated in future work if some conditioning concerning speech activity and phoneme information would be added to the score model.

In addition, we explored different sampling strategies to solve the reverse process at test time which allows us to balance the performance and computational speed of the proposed method. Future work could include other sampling techniques to further reduce the number of diffusion steps \cite{watson2021learning} and thus the computational complexity.

\bibliographystyle{IEEEtran}
\bibliography{bib_clean, static_refs}

\end{document}